\documentstyle{amsppt}
\magnification\magstep1
\TagsOnRight

\def\picture #1 by #2 (#3){\vbox to #2{
       \hrule width #1 height 0pt depth 0pt
       \vfill \special{picture #3}}}

\font\refabs=cmr9
\def\hbar{\mathchar '26\mkern -9muh}   
\def\k#1{#1^{(k)}}
\define\vk#1{#1^{(\vec k)}}
\def\ket#1{|#1\rangle}
\define\bra{\langle}
\define\1#1{\left( \frac\omega{\pi\hbar} \right)^{#1}}
\def\F#1{#1^{\phantom B}_{\Cal F}}
\define\esp#1{\langle#1\rangle}
\define\dis{\k{\Delta_{(q,p)}}}
\define\fo{{\Cal F}}
\define\a{\alpha}
\define\e{\epsilon}
\define\p{\psi}
\define\f{\varphi}
\define\om{\omega}
\define\Om{\Omega}
\define\th{\theta}
\define\R{{\Bbb R}}                    
\define\Z{{\Bbb Z}}                    
\define\Tn{{\Bbb T^n}}                 
\define\TTn{(\Bbb T^n)'}               

\centerline{\bf COHERENT STATES ON THE CIRCLE}
\vskip 1.0cm

\centerline{\bf Jos\'e A. Gonz\'alez and  Mariano A. del Olmo}
\vskip 0.5cm

\centerline{\sl Dpto. de F\'{\i}sica Te\'orica, Universidad de Valladolid,}

\centerline{\sl E--47011 Valladolid, Spain.}

\centerline{\sl E. mail: jagonzalez$\@$fta.uva.es~, olmo$\@$fta.uva.es}

\vskip 1.5cm
\centerline{7 September, 1998}
\vskip 1.5cm

\centerline{\bf Abstract} 
{\refabs 
A careful study of the physical properties of a family of coherent states on the
circle, introduced some years ago by de Bi\`evre and Gonz\'alez in [DG~92], is
carried out. They were obtained from the  Weyl--Heisenberg coherent
states in $L^2(\R)$ by means of the  Weil--Brezin--Zak transformation, they
are labeled by the points of the cylinder $S^1 \times \R$, and they provide a
realization of $L^2(S^1)$ by entire functions (similar to the well-known
Fock--Bargmann construction). In particular, we compute the expectation values of
the position and momentum operators on the circle and we discuss the Heisenberg
uncertainty relation.}

\vfill\eject


\heading 1. Introduction \endheading

This paper is devoted to the study of the physical properties of a family
of coherent states (CS) defined on the circle (i.e., belonging to $L^2(S^1)$) and
labeled by the points of the cylinder. These CS were introduced  by de Bi\`evre
and Gonz\'alez in Ref. [DG~92] and [DG~93], where they were simply sketched.
Therefore, we make here a deep study about them. Our aim
is to contribute to the development of the quantum theory on periodic phase spaces.
Among these phase spaces we pay special attention to the cylinder because of their
relation with physical systems with periodic motion and  their nontrivial topology,
moreover the quantum formalism on the cylinder is far to be completely understood.

On the other hand, it has been proved that families of CS are relevant in the
study of many quantum systems [KS~85] [PE~86], however this formalism 
presents some troubles when one wants to apply it to the cylinder. For
instance, the  cylinder can be seen as a coadjoint orbit of the Euclidean
group of the plane but, in strict sense, the Perelomov method [PE~86] for
constructing CS with this group does not work (de Bi\`evre [DB~89] and Isham
and Klauder [IK~91] have showed two different ways to avoid this problem).
Nevertheless, the CS here introduced have not been obtained by any of these
procedures, but by decomposition of the standard Weyl--Heisenberg CS on $\R$.
The machinery to carry out such a decomposition is the Weil--Brezin--Zak
(WBZ) transform [JA~82] [FO~89], which was originally used for the study of
periodic potentials [ZA~68] [RS~78]. This WBZ transform relates the quantum
formalisms on the plane and on the cylinder (or on the torus as phase space
[BB~96]). Roughly speaking, this procedure maps functions of one variable on
quasiperiodic functions of two variables by a generalization of the Bloch
functions. 

As an application, these CS can be used to study a quantum particle on the
circle as it has been made recently by Kowalski {\sl et al} [KR~96]. Although 
they assume to use CS different from ours and to have obtained a better
approach to this problem, it is easy to prove that their CS are a particular
case of the CS used here, which shows the wider generality of our approach.

The paper is organized as follows. In Section 2 we review the 
main properties of the WBZ transform that plays a central role in our work. 
Section 3 is devoted to the CS on the circle, which are obtained by
decomposition of the standard Weyl--Heisenberg CS on $\R$ (i.e., CS belonging
to $L^2(\R)$); in other words, the CS on the circle are the image of the
Weyl--Heisenberg CS by the WBZ transform. These CS on the circle provide a
realization of the space $L^2(S^1)$ in terms of entire functions as it is
shown in Section 4, in analogy with the Fock--Bargmann representation of
$L^2(\R)$ provided by the Weyl--Heisenberg CS. A part of the results
of Sections 3 and 4 has been published in [DG~92]. Section 5 presents a
generalization of the CS on the circle to a $n$--dimensional torus $\Tn$,
thus we will obtain a family  of CS in $L^2(\Tn)$. The physical properties
are studied in Section 6, paying special attention to the expectation values
of the position and the (angular) momentum operators, and to the Heisenberg
uncertainty relation. The last section is devoted to prove that the CS of
Ref. [KR~96] agree with our CS for a particular value of the parameters that
characterize the latter ones, and to present some conclusions.


\heading 2. The Weil-Brezin-Zak transform \endheading

It is a well known fact that $L^2(\R)$ is isomorphic to $L^2(S^1 \times
S^{1*})$, where $S^{1*}$ is the dual space of $S^1$. This result has been 
used, for instance, in solid state theory to construct the Bloch functions
[ZA~68] [RS~78], as well as for quantum description of periodic variables
[ZA~69]. In this context, we call Weil-Brezin-Zak (WBZ) transform $T$ to the
unitary map from $L^2(\R)$ to $L^2(S^1 \times S^{1*})$ [JA~82] [FO~89]. If we
identify $S^1$ with the interval $[0,a)$ and $S^{1*}$ with $[0, 2\pi/a)$,
then $T$ is explicitly given by
$$
(T\p)(q,k) = \sum_{n=-\infty}^\infty e^{inak} \p(q-na),  \tag2.1
$$
for $\p \in L^2(\R),\ q \in S^1$ and $k \in S^{1*}$. Conversely,
$$
\p(q-na) = \frac a{2\pi} \int_0^{2\pi/a}dk\, e^{-inak}\, (T\p)(q,k),
\qquad  q \in S^1,\ n \in \Z.       \tag2.2
$$
In this way, the functions $T\p$ are periodic in $k$ and quasiperiodic in
$q$,
$$
(T\p)(q+na, k+m\frac{2\pi}a) = e^{inak}\, (T\p)(q,k), \qquad  n,m \in \Z.    
\tag2.3
$$
Note that if we fix a value of $k$ (as we are going to do from now on) the
operator given by (2.1) is a projection onto $L^2(S^1)$, which we will denote
by $\k T$, and we get the so-called constant fiber direct integral
decomposition [RS~78],
$$
L^2(\R) \cong \int_{S^{1*}}^\oplus dk\, L^2(S^1). \tag2.4
$$
In this case, we will frequently use the notation $\k T \p = \k\p$,
and we will say that $\k\p$ is obtained by decomposition of $\p$.


\heading 3. Coherent states on the circle\endheading

In this section we are going to show that a family of CS in $L^2(S^1)$ can 
be constructed by decomposition of the standard Weyl--Heisenberg CS in
$L^2(\R)$. The latter are given as an orbit under the Weyl--Heisenberg group
[KS~85] [PE~86]:
$$
\eta_{y,p}(x) := \exp \bigl(\frac i\hbar(pQ - yP)\bigr)\, \eta_0(x) =
\exp\bigl(\frac i\hbar p(x - \frac y2)\bigr)\, \eta_0(x-y),   \tag3.1
$$
where $x,y,p \in \R$ and $\eta_0 \in L^2(\R)$ is a fiducial state,
which is usually chosen to be a normalized gaussian:
$$
\eta_0(x) = \1{1/4} \exp (-\frac\om{2\hbar} x^2).    \tag3.2
$$
Now, we can use (2.1) to construct the functions $\k\eta_{y,p} \in L^2(S^1)$
(or $\ket{y,p;k}$ in Dirac's notation), and it is natural to ask if, for each
value of $k$, this set of functions will be also a set of CS, labeled by
suitable values of $y$ and $p$. The answer is positive, according to the
generalized definition of CS given in [KS~85]: simply, a family of states 
depending continuously on a set of labels and fulfilling a resolution of
the unity. They are not constructed by Perelomov's method [PE~86], as an orbit
under  a Lie group representation. Actually, we have here a non--trivial
example of the ``reproducing triples'' introduced in [AA~91].

\proclaim{(3.1) Theorem}
For each $k \in S^{1*}$, the family $\bigl\{\k\eta_{q,p}\equiv \ket{q,p;k}\
\bigm|\ (q,p)  \in S^1\times\R\bigr\}$, where $\eta_{q,p}$ is given by (3.1)
with $\|\eta_0\| = 1$, is a set of coherent states in $L^2(S^1)$; i.e.,
they verify the following resolution of unity: 
$$
\frac1{2\pi\hbar} \int_0^a dq\int_{-\infty}^\infty dp\, \ket{q,p;k}
\langle q,p;k| =  I.\tag3.3
$$
\endproclaim

The proof consists of a simple calculation, using the definitions (2.1) and
(3.1) [GO~96]. If we choose $\eta_0$ according to (3.2), these CS take the 
form 
$$
\multline \k\eta_{q,p}(q') = \1{1/4} 
\exp {\left(\frac i{2\om\hbar} pz^*\right)}\,
\exp {\left(-\frac1{2\om\hbar}(z^* - \om q')^2\right)}  \\
\times \th\!\left(i\frac a{2\hbar}(z^*-\om q'-ik\hbar); 
\rho_1 \right)\!, \endmultline    \tag3.4
$$
where $z^* = \om q + ip$, $\displaystyle \rho_1 =
\exp \left(-\frac{a^2\om}{2\hbar}\right)$, and $\th(z;\rho) = 
\sum\limits_{n=-\infty}^\infty \rho^{n^2} e^{2inz}$, $|\rho|<1$, is the Theta
function (sometimes denoted by $\th_3$) [WW~27] [AS~72] [ER~81] [MU~83].

As a corollary to the preceding theorem, we present a typical property of every
set of CS [KS~85].

\proclaim{(3.2) Corollary}
The mapping $\k W \:  L^2(S^1) \to L^2(S^1\times\R)$ given by
$$
(\k W \varphi)(q,p) = \langle q,p;k \ket\varphi    \tag3.5
$$
is an isometry, and $\k W\bigl(L^2(S^1)\bigr)$ is a reproducing kernel space,
with kernel
$$
\frac1{2\pi\hbar} \langle q',p';k\ket{q,p;k}.
$$
\endproclaim

To compute this kernel, let us consider the orthonormal basis in $L^2(S^1)$:
$$
\biggl\{ \ket{n;k} \equiv \frac1{\sqrt a} \exp i(\frac{2\pi}a n + k)q, \qquad
n \in \Z, \ k \in S^{1*} \ \text{fixed} \biggr\}.\tag3.6
$$
Then we can write
$$
\ket{q,p;k} = \sum_{n=-\infty}^\infty c_n^{(q,p;k)} \ket{n;k},\tag3.7
$$
where the coefficients are 
$$
c_n^{(q,p;k)} = \sqrt{\frac{2\pi\hbar}a}\, e^{i[p/(2\hbar)-(2\pi n/a+k)]q}\,
\widetilde\eta_0\bigl((\frac{2\pi}a n + k)\hbar - p\bigr),    \tag3.8
$$
$\widetilde\eta_0$ being the Fourier transform of $\eta_0$. Now, using (3.2)
and (3.8), we easily obtain
$$
\align \langle q',p';k\ket{q,p;k} &=
\sum_{n=-\infty}^\infty \langle q',p';k\ket{n;k} \langle n;k\ket{q,p;k}
= \sum_{n=-\infty}^\infty c_n^{(q',p';k)*}c_n^{(q,p;k)}  \\
&= \frac2a \sqrt{\frac{\pi\hbar}\om} e^{ik(q'-q)}
e^{i(qp-q'p')/2\hbar} e^{-[(\hbar k-p)^2 + (\hbar k-p')^2]/2\om\hbar} \\
&\qquad\qquad \times \th\!\left(\frac\pi a
[(q'-q) + \frac i\om (2\hbar k - p - p')]; \rho_2\right)\!,  \tag3.9
\endalign
$$
where $\displaystyle \rho_2 = \exp \left(-\frac{4\pi^2\hbar}{\omega a^2}\right)$.

By the way, we see that these CS are not normalized. It follows immediately from
(3.9) that
$$
\bra q,p;k\ket{q,p;k} = \frac2a \sqrt{\frac{\pi\hbar}\om}\,
e^{-(\hbar k - p)^2/(\om\hbar)}\  
\th\!\left(i\frac{2\pi}{\om a} (\hbar k - p); \rho_2 \right)\!.   \tag3.10
$$
Taking into account the identity
$$
\th(z; \rho_2) = \frac a2 \sqrt{\frac\om{\pi\hbar}}\,
e^{-\om a^2 z^2/(4\pi^2\hbar)}\,
\th\!\left(-i\frac{\om a^2}{4\pi\hbar}z;\, \rho_1^{1/2} \right)\!, 
\tag3.11
$$
which is easily deduced from the so-called functional equation of $\th$ 
[ER~81] [MU~83], we obtain as well the expression
$$
\bra q,p;k\ket{q,p;k} = \th\!\left(\frac a{2\hbar} (\hbar k - p);
\rho_1^{1/2} \right)\!.    \tag3.12
$$


\head 4. A realization of $L^2(S^1)$ by analytic functions\endhead

Let us consider again the isometry $\k W$ given by (3.5). If we define the
new mapping
$$
\align
(\k B\varphi)(z) &= \exp(\frac i{2\om\hbar} pz)\, (\k W\varphi)(q,p)  
\tag4.1a \\ 
&= \exp(\frac i{2\om\hbar} pz)\, \langle q,p;k \ket\varphi,
\qquad \varphi \in L^2(S^1),  \tag4.1b 
\endalign
$$
with $z = \om q - ip$, then $\k B\varphi$ is an analytic function on 
$S^1 + i\R$ (because $\th(z;\rho)$ is an entire function of $z$).
This suggests us to search for a representation of $L^2(S^1)$ by entire
functions, similar to the standard Fock--Bargmann representation of
$L^2(\R)$ [BA~61] [PE~86] [FO~89]. In this context, it is quite natural to
define a new set of CS, labeled by $z^* = \om q + ip$, by
$$
\align
\ket{z^*; k} &= \exp(-\frac i{2\om\hbar} pz^*)\, \ket{q,p;k}, \tag4.2a  \\
\k\eta_{z^*}(q') &= \1{1/4} \exp\left(-\frac{(z^*-\om q')^2}
{2\om\hbar}\right) \th\!\left(i\frac a{2\hbar}(z^*-\om q'-ik\hbar);
\rho_1\right)  \tag4.2b 
\endalign
$$
such that we simply have
$$
(\k B\varphi)(z) = \langle z; k \ket\varphi, \qquad
\varphi \in L^2(S^1).   \tag4.3
$$
Note that we write $\ket{z^*; k}^\dagger = \langle z; k|$. Since 
$\ket{z^* + \om a; k} = e^{-iak}\ket{z^*; k}$, as it is easy to check, we can 
extend $(\k B\varphi)(z)$ to the whole of $\Bbb C$, getting so an entire function
of
$z$,
$\forall \varphi \in L^2(S^1)$. Moreover, the CS $\ket{z^*; k}$ fulfill the
resolution of the unity:
$$
{1 \over 2\pi\hbar}\int_0^a dq\int_{-\infty}^\infty dp\,
e^{-p^2/(\om\hbar)} \ket{z^*; k} \langle z; k| = I.\tag4.4
$$
Hence $\k B$ is an isometry from $L^2(S^1)$ into the space
$$
\multline \fo = \biggl\{\, \p(z) \text{ entire, } \p(z + \om a) =
e^{iak}\, \p(z) \text{ and }     \\     \|\p\|_{\fo}^2 = \frac1{2\pi\hbar}
\int_0^a dq\int_{-\infty}^\infty dp\, e^{-p^2/(\om\hbar)}\, 
|\p(z)|^2 < \infty, \ z = \om q - ip \,\biggr\}.  \endmultline\tag4.5
$$

We see that the space $\fo$ is similar to the usual Fock space. Since $\k B$ 
also maps $L^2(S^1)$ onto $\fo$, as we will see, we have a complete analogy
with the standard case. Obviously, we can define the following orthonormal
set in $\fo$: $$
\bigl\{\, \p_n(z) := \bigl(\k B\ket{n;k}\bigr)(z) = \langle z;k \ket{n;k}\,
\ |\,\ n \in \Z \,\bigr\},\tag4.6
$$
and it is not hard to compute the functions
$$
\p_n(z) = \left(\frac{4\pi\hbar}{a^2\om}\right)^{1/4}
\exp\bigl(-\frac{\hbar}{2\om} (\frac{2\pi}a n + k)^2\bigr)\,
\exp\bigl(\frac i\om (\frac{2\pi}a n + k)z\bigr). \tag4.7
$$
To prove that $\k B$ is onto is equivalent to prove that these functions 
form a basis for $\fo$. But, after (4.7), this amounts to the existence of a
Fourier series for any $\p \in \fo$, as it is really the case, i.e., 
$$
\p(z) = \sum_{n=-\infty}^\infty a_n\, e^{i(2\pi n/a + k)z/\om},
\qquad \forall \p \in \fo,  \tag4.8
$$
because of the quasiperiodicity of the functions in $\fo$ (we have introduced,
for convenience, a factor $e^{ikz/\om}$ in the usual Fourier series). Using
(4.7) and the orthonormality of the set $\{\, \p_n \,\}$, the expression 
(4.8) becomes
$$
\p(z) = \sum_{n=-\infty}^\infty \F{(\p_n|\p)}\, \p_n(z), \qquad
\forall \p \in \fo,  \tag4.9
$$
where $\F{(\cdot\,|\,\cdot)}$ denotes the inner product of $\fo$ and
$$
\F{(\p_n|\p)} = a_n \left(\frac{a^2\om}{4\pi\hbar}\right)^{1/4}
e^{\hbar(2\pi n/a + k)^2/(2\om)}.     \tag4.10
$$
Clearly, there is a one-to-one correspondence between the coefficients
$\F{(\p_n|\p)}$ and $a_n$, so the set $\{\, \p_n \,\}$ is a basis and $\k B$ 
is onto. 

We can write now some expressions for the inverse $B^{-1}$ of $\k B$:
$$
\align 
\ket{B^{-1}\p} &= \sum_{n=-\infty}^\infty
\F{(\p_n|\p)}\, \ket{n;k}    \tag4.11a \\
&= \frac1{2\pi\hbar} \int_0^a dq \int_{-\infty}^\infty dp\,
e^{-p^2/(\om\hbar)}\, \p(z)\, \ket{z^*;k}, \qquad z = \om q - ip.   
\tag4.11b    \endalign
$$


\heading 5. Coherent states on the torus\endheading

All the results of the precedent sections can be easily generalized to a
higher number of dimensions. To this purpose, let us take a unitary basis
$\{\, \vec e_1, \vec e_2, \ldots, \vec e_n \,\}$ in $\R^n$ as well as a set 
of real numbers $\{\, a_1, a_2, \ldots, a_n \,\}$ and let us consider the
associated lattice $\Cal L$ [RS~78], that is,
$$
\Cal L = \{\, \vec a \in \R^n \,\ |\,\ \vec a = 
\sum_{i=1}^n m_i a_i \vec e_i, \  m_i \in \Z \,\}.   \tag5.1
$$
In the same way, we define the dual basis 
$\{\, \vec\e_1,\ldots,\vec\e_n \,\}$ by 
$\vec\e_i \cdot \vec e_j = \delta_{ij}$ and the dual lattice by 
$$
\Cal L' = \{\, \vec b \in \R^n \,\ |\,\ \vec b = \sum_{i=1}^n m_i
\dfrac{2\pi}{a_i} \vec\e_i, \  m_i \in \Z \,\}.  \tag5.2
$$
The corresponding basic cells $\Tn$ and $\TTn$ are $n$--dimensional torus,
$$
\align \Tn &= \{\, \vec q \in \R^n \ |\ \vec q = \sum_{i=1}^n q_i \vec e_i,
\ 0 \leq q_i < a_i \,\},    \tag5.3  \\
\TTn &= \{\, \vec k \in \R^n \,\ |\,\ \vec k = \sum_{i=1}^n
k_i \vec\e_i, \ 0 \leq k_i < \frac{2\pi}{a_i} \,\}.   \tag5.4  \endalign
$$

We shall define the n-dimensional Weil-Brezin-Zak transform $T$ as a unitary 
map from $L^2(\R^n)$ to $L^2\bigl(\Tn \times \TTn\bigr)$ [JA~82] [FO~89],
given by 
$$
(T\p)(\vec q,\vec k) = \sum_{\vec a \in \Cal L} e^{i\vec a \cdot \vec k}
\p(\vec q - \vec a),  \qquad  \vec z \in \Tn, \ \vec w \in \TTn,  \tag5.5
$$
with $\p \in L^2(\R^n)$. The functions $T\p$ verify
$$
(T\p)(\vec q + \vec a,\vec k + \vec b) = e^{i\vec a\cdot\vec k}
(T\p)(\vec q,\vec k), \qquad   \vec a \in \Cal L, \ \vec b \in \Cal L'.
\tag5.6 
$$
From now on, we are going to fix a value of $\vec k$, so that the expression
(5.5) defines a projection $\vk T$ onto $L^2(\Tn)$. We will use the notation
$\vk T\p = \vk\p$.

Coherent states on the torus are obtained as the image under $\vk T$ of the
$n$--dimensional Weyl--Heisenberg CS $\eta_{\vec q,\vec p} \in L^2(\R^n)$,
which we write as
$$
\eta_{\vec q,\vec p}(\vec x) = \exp\bigl(\frac i\hbar \vec p \cdot
(\vec x - \frac {\vec q}2)\bigr)\, \eta_0(\vec x - \vec q\,),  \tag5.7
$$
where $\vec x, \vec q, \vec p \in \R^n$, and the fiducial state $\eta_0 \in
L^2(\R^n)$ is chosen to be a normalized gaussian:
$$
\eta_0(\vec x) =  \1{n/4} \exp(-\frac\om{2\hbar} {\vec x}^2).  \tag5.8
$$
In this case, the functions $\vk\eta_{\vec q,\vec p}$ take the form [GO~96]
$$
\multline \vk\eta_{\vec q,\vec p}({\vec q\,'}) =
\1{n/4} e^{-i\vec p\cdot\vec q/(2\hbar)} e^{i\vec p\cdot{\vec q\,'}/\hbar}
e^{-\om(\vec q-{\vec q\,'})^2/(2\hbar)}         \\
\times \varTheta\!\left(\frac1{2\hbar} \Delta [\hbar \vec k - \vec p +
i\om G (\vec q - \vec q\,')] \biggm| \Om\right)\!,    \endmultline \tag5.9
$$
where $G$ is the $n \times n$ symmetric matrix of the lattice, whose elements
are $g_{ij} = \vec e_i \cdot \vec e_j$; $\Delta$ is a $n \times n$ diagonal
matrix with elements $a_i$; $\Om$ is another $n \times n$ matrix given by
$$
\Om = i \frac\om{2\pi\hbar}\, \Delta G \Delta;   \tag5.10
$$
and $\varTheta$ is the n--dimensional Theta function [MU~83]:
$$
\varTheta(\vec z \,|\Om) = \sum_{\vec m \in \Z^n}
\exp(i\pi \vec m \cdot\Om\vec m)\, \exp(2i \vec m \cdot\vec z).  \tag5.11
$$
We have thus the following n--dimensional version of Theorem (3.1):

\proclaim{(5.1) Theorem}
For each $\vec k \in \TTn$, the family of functions
$$
\bigl\{\, \vk\eta_{\vec q,\vec p} \equiv \ket{\vec q,\vec p; \vec k} \ 
\bigm| \  (\vec q,\vec p) \in \Tn \times \R^n \,\bigr\}, \tag5.12
$$
given by (5.9), is a set of coherent states in $L^2(\Tn)$; 
i.e., they verify the resolution of unity:
$$
\frac1{(2\pi\hbar)^n} \int_{\Tn} d\vec q \int_{R^n} d\vec p\,
\ket{\vec q,\vec p;\vec k} \bra \vec q,\vec p;\vec k| = I.\tag5.13
$$
\endproclaim

Most generalizations of the one--dimensional results are straightforward
[GO~96]. We simply are going to write here the expression for the product
$\bra {\vec q\,'},{\vec p\,'};\vec k \ket{\vec q,\vec p;\vec k}$. After a
rather lengthy calculation, we obtain
$$
\multline \bra {\vec q\,'},{\vec p\,'};\vec k \ket{\vec q,\vec p;\vec k} \\
= \frac{2^n}{\sqrt g A} \left(\frac{\pi\hbar}{\om}\right)^{n/2}
e^{i(\vec p\cdot\vec q - {\vec p\,'}\cdot{\vec q\,'})/(2\hbar)}
e^{i\vec k\cdot({\vec q\,'} - \vec q)} e^{-[(\hbar\vec k-{\vec p\,'})^2 +
(\hbar\vec k-\vec p)^2]/(2\om\hbar)}    \\    \times
\varTheta\!\left( \pi\Delta^{-1}\bigl[{\vec q\,'} - \vec q + \frac i\om
G^{-1} (2\hbar\vec k - \vec p-{\vec p\,'})\bigr] \biggm| \Om'\right)\!,  
\endmultline    \tag5.14
$$
where $g = \det G$, $A = a_1 a_2 \cdots a_n$ and $\Om' = -2\Om^{-1}$.
Therefore, we get also
$$
\bra\vec q,\vec p;\vec k\ket{\vec q,\vec p;\vec k} = \frac{2^n}{\sqrt g A}
\left(\frac{\pi\hbar}{\om}\right)^{n/2}
e^{-(\hbar\vec k-\vec p\,)^2/(\om\hbar)}\,
\varTheta\!\left(i\frac{2\pi}\om \Delta^{-1} G^{-1} (\hbar\vec k-\vec p\,)
\biggm| \Om' \right)\!.      \tag5.15
$$
Finally, it can be shown [GO~96] that when the lattice is orthogonal, the CS
$\vk\eta_{\vec q,\vec p}$ factorize out like a product of one--dimensional CS
given by (3.4), i.e.,

$$
\vk\eta_{\vec q,\vec p}({\vec q\,'}) = \prod_{i=1}^n
\eta^{(k_i)}_{q_i,p_i}(q'_i).  \tag5.16
$$


\heading 6. Physical properties of the CS on the circle\endheading

This section is devoted to discuss the physical properties of the
CS on the circle introduced in Section 3 (a complete study has been realized 
in [GO~96]). As these states have been constructed by decomposition of the
standard Weyl--Heisenberg CS in $L^2(\R)$, it seems to us that comparison
between both cases could be illustrative. Moreover, it is known that the
Weyl--Heisenberg CS have very nice quasiclassical properties, for instance, 
to minimalize the Heisenberg uncertainty relation, and it would be of great
interest to reproduce such behaviour on the circle. As a matter of fact, we 
are going to see that the physical properties of the CS on the circle depend
mainly on some dimensionless parameter, related to the spread of the initial
standard CS. If this spread was smaller than the length $a$ of the circle, we
get CS on the circle very similar to the standard CS. But if such spread was
comparable or bigger than $a$, the CS on the circle are rather like plane
waves.

We will discuss also the relation between the CS parameters $q$, $p$ and the
expectation values in these states of the position and momentum operators on
$L^2(S^1)$. Whereas for standard CS in $L^2(\R)$ both things are
the same, this is not the case on the circle. So, firstly we will recall
the correct definitions for the position and momentum operators on $L^2(S^1)$
(which show some significant differences from their analogous on the real
line). Then, we will compute the expectation values of these operators and,
finally, we will devote some attention to the Heisenberg uncertainty relation
on the circle, but in a different and more suitable form of the usual one
on the real line.

In order to provide an easier understanding of the somewhat complicated
expressions, we are going to illustrate our results with several figures. 
In any case, it has been possible to realize a complete analytic study
[GO~96].


\subhead 6.1 Quantum mechanics on the circle\endsubhead

\smallskip
The topology of the circle has important consequences for the quantum
formalism on this configuration space. Indeed, experience shows that a direct
translation of the formalism on the real line leads to serious inconsistencies
[CN~68] [ZA~69] [LE~76]. For instance, it is known that the (angular) momentum
operator on $L^2(S^1)$ has discrete spectrum. Moreover, functions in its 
domain have to verify the constraint $\f(a) = e^{iak} \f(0)$, where $a$ is the
length of the circle and $k \in [0, 2\pi/a)$ is a parameter as in Section 2
[RS~75]. Thus, in fact, there is not one but a family of momentum operators on
$L^2(S^1)$, labeled by $k$ and which we denote by $\k P$. As a consequence, a
canonical commutation relation as in $L^2(\R)$
$$
[Q,\k P] = i\hbar, \tag6.1
$$
with position operator $Q$ defined as usual, is inconsistent in $L^2(S^1)$.
Heisenberg's uncertainty relation is even more troublesome, because of the
compact spectrum of $Q$ on $L^2(S^1)$. In effect, this relation allows the
position dispersion to be bigger than $a$, which has no physical meaning.

All these problems can be solved choosing the unitary operator $E =
\exp(i2\pi Q/a)$ as a better representation for the position on the circle
[LE~76]. It has precisely the circle as spectrum and its commutator with the
momentum operator is
$$
[\k P, E] = \frac{2\pi\hbar}a\, E,    \tag6.2
$$
which poses no domain problems. From this fundamental relation (6.2) we can 
also deduce an uncertainty relation more suitable for the circle [LE~76].
Since $E$ is unitary but not selfadjoint, the dispersion $\Delta\, E$ should
be defined in the form
$$
(\Delta\, E)^2 :=  \esp{E^\dagger E} - |\esp E|^2 = 1 - |\esp E|^2,  \tag6.3
$$
so that relation (6.2) yields, by the usual method, the following Heisenberg
uncertainty relation:
$$
(\Delta\, \k P)^2\, \frac{(\Delta\, E)^2}{1 - (\Delta\, E)^2}
\ge \left(\frac\pi a \hbar\right)^2.   \tag6.4
$$
Note that now, when $\Delta\, \k P = 0$ we must have $\Delta\, E = 1$, which
is a more appropriate result because of the compactness of the position
variable on the circle. Moreover, relation (6.4) reduces to the usual
Heisenberg uncertainty relation when $\Delta\, E \ll 1$ [LE~76]. Thus, we
will call $E$ the ``angle'' operator and from now on we will use it as the
quantum representation for the position on the circle.


\subhead 6.2 Physical properties of the CS on the circle\endsubhead

\smallskip
\noindent 6.2.1 {\underbar{\sl Probability density}.} \ Let us begin the 
study of the  basic physical properties of the CS $\ket{q,p;k}$ by computing
its probability density $\Cal P_{q,p;k}(q')$. The wave function
$\k\eta_{q,p}(q')$ of these states is given by expression (3.4). As they are
not normalized, the probability density will be
$$
\Cal P_{q,p;k}(q') = \frac{|\k\eta_{q,p}(q')|^2}{\bra{q,p;k}\ket{q,p;k}},
\tag6.5
$$
that, making use of the expression (3.12), yields
$$
\Cal P_{q,p;k}(q') = \1{1/2} e^{-\om (q-q')^2/\hbar}
\frac{\big|\th\big(a[(k\hbar-p) + i\om(q-q')]/(2\hbar); \rho_1\big)\big|^2}
{\th\big(a(k\hbar-p)/(2\hbar); \rho_1^{1/2}\big)}.   \tag6.6
$$

\topinsert

\centerline{\picture 130.6mm by 173.8mm (figure1)}

\botcaption{Figure 1} The functions $a\, \Cal P_\a(u-\frac12,0)$ (left) and
$a\, \Cal P_\a(u-\frac12,\tfrac12)$ (right), for several values of $\a$.
\endcaption

\bigskip\bigskip

\endinsert

In order to clarify the notation, we are going to introduce two new variables,
$$
u := \frac1a(q' - q),\qquad\qquad  v:= \frac a{2\pi\hbar}(p - k\hbar),\tag6.7
$$
as well as the dimensionless parameter
$$ 
\a := \frac{a^2}{2\hbar}\, \om. \tag6.8
$$
In this way, the probability density, from now on denoted by $\Cal P_\a(u;v)$,
looks like
$$
\Cal P_\a(u;v) = \frac1a \sqrt{\frac{2\a}\pi} e^{-2\a u^2}\,
\frac{|\th(\pi v + i\a u; e^{-\a})|^2}{\th(\pi v; e^{-\a/2})}.  \tag6.9
$$
This is a periodic function of $u$ and $v$, in both cases with period 1. This
corresponds to a period $a$ for $q'-q$ and a period $2\pi\hbar/a$ for $p$.
To give a general idea of its main properties, we show in Figure 1 some 
significant cases. We observe that for high values of $\a$ (approximately 
$\a >15$) the probability density is, with a good accuracy, a Gaussian
regardless of the value of $v$. That is, we have the same result as for the
standard Weyl--Heisenberg CS. Note that, for these values of $\a$, the width
of the Gaussian is always smaller than $a$. On the other hand, for small
values of $\a$ the probability density is no longer a Gaussian and its shape
depends crucially on the value of $v$. Only when $v = 1/2$ (i.e., 
$p = \bigl((2n+1)\pi/a + k\bigr)\hbar$, with $n \in \Z$), it looks always 
as a ``wave packet'' for all the values of $\a$ (right side of the Figure 1).
In all the other cases it tends to be a plane wave when $\a \to 0$ (in the
left side of the Figure 1 we show the case $v = 0$)

\bigskip


\noindent
6.2.2 {\underbar{\sl Expectation value of the angle operator}.} \ To compute 
the expectation value of $E$ in the CS $\ket{q,p;k}$, we will make use of the
following relation
$$
E\ket{q,p;k} = e^{i\pi q/a} \ket{q,p+\frac{2\pi}a\hbar;k}, \tag6.10
$$
which is easily deduced from the obvious action of $E$ on the orthonormal
basis $\ket{n;k}$ in $L^2(S^1)$ (see expression (3.6)),
$$
E \ket{n;k} = \ket{n+1;k},  \qquad \forall n \in \Z,    \tag6.11
$$
as well as from the expression (3.8) for the coefficients of the CS
$\ket{q,p;k}$ in this basis. We will denote the expectation value of $E$ by
$\esp E(u,v)$, with $v$ as in (6.7) but
$$
u := \frac qa, \tag6.12
$$
from now on. We also continue using the parameter $\a$ defined in (6.8). 
Taking together the equations (6.10), (3.9), (3.11) and (3.12) we finally
arrive at 
$$
\align
\esp E(u,v) &= \frac{\bra q,p;k|E\ket{q,p;k}}{\bra{q,p;k}\ket{q,p;k}} \\
&= e^{i2\pi u} e^{-\pi^2/(2\a)}\, \frac{\th\bigl(\pi(v-\frac12);
e^{-\a/2}\bigr)} {\th(\pi v; e^{-\a/2})}.    \tag6.13    \endalign
$$
Of course, this is a periodic function of $u$ but also an even periodic 
function of $v$ with period 1. As all the factors excepting $e^{i2\pi u}$ 
are real positive [WW~27] [ER~81] [MU~83], we can write
$$
\esp E(u,v) = e^{i2\pi u}\, |\esp E|(v).\tag6.14
$$

\topinsert

\centerline{\picture 130.6mm by 130.6mm (figure2)}

\botcaption{Figure 2} The function $|\esp E|(v)$, for several values of $\a$.
\endcaption

\bigskip

\endinsert

We show the function $|\esp E|(v)$ in Figure 2, for some values of $\a$. 
Note that, in general, it is not possible to interpret the expectation value 
of $E$ as a measure of the average position of the CS  on the circle because
of the dependence on $v$. However, observe in Figure 2 that for high values 
of $\a$ the function $|\esp E|(v)$ is almost constant, so in these cases we
would get the usual interpretation of the CS parameter $q$ as the average
position of the quantum state. On the other hand, we have seen in Figure~1
that for small values of $\a$ most of the CS are nearly plane waves, hence it
is not so important that the average position in these states cannot be well
defined.

\bigskip

\noindent
6.2.3 {\underbar{\sl Expectation value of the momentum operator}.} 
\ We begin the calculation observing that the vectors of the basis 
$\ket{n;k}$ in  $L^2(S^1)$ (see expression (3.6)) are eigenvectors of the
momentum operator $\k P$, 
$$
\k P \ket{n;k} = \hbar (\frac{2\pi}a n + k)\, \ket{n;k}.  \tag6.15
$$
Thus, we can write
$$
\bra q,p;k| \k P \ket{q,p;k} = \hbar \sum_{n=-\infty}^\infty
(\frac{2\pi}a n + k)\, \bigl| c_n^{(q,p;k)}\bigr|^2,    \tag6.16
$$
where the coefficients $c_n^{(q,p;k)}$ are given by the expressions (3.8) and
(3.2). Using also the formulae (3.11) and (3.12) we finally find
$$
\esp{\k P}(p) =
\frac{\bra q,p;k|\k P\ket{q,p;k}}{\bra{q,p;k}\ket{q,p;k}}    
= p + \frac{\hbar\a}{2a} \frac{\th'(\pi v;e^{-\a/2})}{\th(\pi v; e^{-\a/2})},
\tag6.17  
$$
where, for the sake of clarity, we use the two variables $p$, $v$ at the same
time, and
$$
\th'(z;\rho) = \frac d{dz} \th(z;\rho) =
2i \sum_{n=-\infty}^\infty n \rho^{n^2} e^{2inz},  \qquad  |\rho|<1. \tag6.18
$$
It is interesting to note that when $v = n/2$, with $n \in \Z$, i.e.
$p = (n\pi/a + k)\hbar$, the expression (6.17) reduces to $\esp{\k P}(p) = p$
as in the standard CS case. For another values of $v$, the difference between
$\esp{\k P}$ and $p$ depends on the parameter $\a$. To show that, let us 
first rewrite the equation (6.17) using only the variable $v$:
$$
\esp{\k P}(v) = \frac{2\pi\hbar}a \left(v + \frac\a{4\pi}
\frac{\th'(\pi v;e^{-\a/2})}{\th(\pi v; e^{-\a/2})}\right) + k\hbar. \tag6.19
$$
Now, we represent the function $(a/{2\pi\hbar}) \left(\esp{\k P}(v) -
k\hbar\right)$ in Figure 3, for some values of $\a$ (remember that 
$2\pi\hbar/a$ is the ``natural unit'' for $p$). We see that for high values
of $\a$, the CS parameter $p$ is a good approximation for the expectation
value of the momentum operator. But for small values of $\a$, this
expectation value tends to take some discrete values for almost all the
values of $v$ [GO~96]. These are the ``plane wave'' states of Figure 1.

\pageinsert

\centerline{\picture 130.6mm by 80.5mm (figure3)}

\vfil

\centerline{\picture 130.6mm by 80.1mm (figure3b)}

\bigskip

\botcaption{Figure 3} The function $\dfrac a{2\pi\hbar} \left(\esp{\k P}(v) -
k\hbar\right)$, for several values of $\a$.
\endcaption

\bigskip

\endinsert

\bigskip


\noindent
6.2.4 {\underbar{\sl Heisenberg uncertainty relation}.} 
\ We conclude the  study of the basic phy\-si\-cal properties of the CS
$\ket{q,p;k}$ with some comments about the Heisenberg uncertainty relation
for these states. In the following we will try to verify if some of the CS
$\ket{q,p;k}$ minimalize relation (6.4), which  has to be used on the circle, 
as we remarked above.

Let us denote by $\dis A$ the dispersion of an operator $A$ in the CS
$\ket{q,p;k}$. We begin computing this dispersion for the angle operator $E$.
According to expression (6.3) we get
$$
(\dis E)^2 = 1 - |\esp E(u,v)|^2 = 1 - |\esp E|(v)^2,   \tag6.20
$$
where $|\esp E|(v)$ can be obtained from expression (6.13).

On the other hand, the dispersion $\dis{\k P}$ of the momentum operator 
requires a few more calculations. Firstly, we have to compute the expectation
value $\esp{(\k P)^2}$, which after (6.15) can be written as
$$
\esp{(\k P)^2}(p) = \frac{\hbar^2} {\bra{q,p;k}\ket{q,p;k}}
\sum_{n=-\infty}^\infty (\frac{2\pi}a n + k)^2\,\bigl| c_n^{(q,p;k)}\bigr|^2.
\tag6.21
$$
Hence, making use again of the formulae (3.8), (3.2), (3.11) and (3.12) we 
get, after a rather lengthy but straightforward calculation,
$$
\esp{(\k P)^2}(p) = \left(\frac{\hbar\a}{2a}\right)^2
\frac{\th''(\pi v; e^{-\a/2})} {\th(\pi v; e^{-\a/2})} + p \left(
\frac{\hbar\a}a \frac{\th'(\pi v;e^{-\a/2})}{\th(\pi v; e^{-\a/2})} + p
\right) + \frac{\hbar^2\a}{a^2},   \tag6.22
$$
with $\th''(z;\rho) = d^2\th(z;\rho)/dz^2$. Finally, equations (6.22) and
(6.17) taken together yield
$$
(\dis{\k P})^2 = \left(\frac\hbar a\right)^2 \left[\frac{\a^2}4
\left(\frac{\th''(\pi v; e^{-\a/2})}{\th(\pi v; e^{-\a/2})} -
\frac{\th'(\pi v;e^{-\a/2})^2}{\th(\pi v; e^{-\a/2})^2}\right) +
\a\right]\!.    \tag6.23
$$

We are now able to discuss the uncertainty relation (6.4) for the CS
$\ket{q,p;k}$. Firstly, we define the uncertainty function
$$\align
\Delta(v) &:= \frac a{2\pi} \frac{\dis E}{\sqrt{1 - (\dis E)^2}}\, \dis{\k P} 
\\ &=  \frac a{2\pi} \left(\frac1{|\esp E|(v)^2} - 1\right)^{1/2} \dis{\k P}.
\tag6.24\endalign
$$
In this way, relation (6.4) reduces to
$$
\Delta(v) \ge \frac\hbar2,   \tag6.25
$$
which looks more similar to standard Heisenberg uncertainty relation, making
thus the present discussion more intuitive. In view of expressions (6.13) and
(6.23) we arrive at the following formula:

\topinsert

\centerline{\picture 130.6mm by 130.6mm (figure4)}

\botcaption{Figure 4} The function $\dfrac 2\hbar\, \Delta(v)$ for several
values of $\a$. \endcaption

\bigskip

\endinsert

$$
\multline \Delta(v)^2 = \left(\frac\hbar{2\pi}\right)^2 \a
\left[ e^{\pi^2/\a}\, \frac{\th(\pi v; e^{-\a/2})^2} {\th\bigl(\pi(v-\frac12);
e^{-\a/2}\bigr)^2} - 1\right]      \\       \times \left[\frac\a4
\left(\frac{\th''(\pi v; e^{-\a/2})} {\th(\pi v; e^{-\a/2})}
- \frac{\th'(\pi v; e^{-\a/2})^2} {\th(\pi v; e^{-\a/2})^2}\right)
+ 1\right]\!.     \endmultline   \tag6.26
$$
We have represented the function $(2/\hbar)\, \Delta(v)$ in Figure 4, for 
some values of $\a$. Observe its somewhat curious appearance. We remark that
variable $v$ is related to the CS parameter $p$, and that parameter $\a$
measures if the CS $\ket{q,p;k}$ is or not similar to a standard CS. In
Figure 4, the value 1 on the vertical scale would correspond to a minimum
uncertainty state, and in fact we see that for high values of $\a$ the
function $\Delta(v)$ tends to this minimum, regardless of the value of $v$
[GO~96]. Nevertheless, none of the CS $\ket{q,p;k}$ is a real minimum
uncertainty state, although we can obtain states so close to this limit as 
we wish, taking $\a$ sufficiently high.

On the contrary, when the value of $\a$ is small we can see that the 
behaviour of the uncertainty relation for the CS $\ket{q,p;k}$ depends on 
the particular value of $p$, that is, $v$. As $\Delta(v)$ is an even periodic
function of $v$, we just need to consider the values $0 \le v \le \frac12$.
Thus, it can be proved [GO~96] that
$$
\lim_{\a\to 0} \Delta(v) = \cases \dfrac{\sqrt2}2 \hbar, &\text{if } v = 0,\ 
\text{i.e., } \ p = (2n \dfrac{\pi}a + k)\hbar, \quad n \in \Z;  \\
\dfrac{\sqrt3}2 \hbar, &\text{if } v = \dfrac12, \text{i.e., }\
p = \bigl( (2n+1)\dfrac{\pi}a + k \bigr)\hbar, \quad n \in \Z;  \\
\hbar, &\text{in any other case.}   \endcases   \tag6.27
$$
In other words, the uncertainty function $\Delta$ is upperly bounded, at
worst, by $\hbar$! Hence, we conclude that for all the family of CS
$\ket{q,p;k}$ we have
$$
\hbar > \Delta(v) > \frac\hbar2,   \tag6.28
$$
that, although strictly speaking does not correspond to minimum uncertainty
states, as a matter of fact shows a quite good behaviour of the CS 
$\ket{q,p;k}$ in this subject. The best behaviour is obtained for those states
associated to the value $v = 0$.


\head 7. Conclusions\endhead

As we mention in the Introduction, a family of CS on the circle has been
introduced in Ref. [KR~96]. These new CS are a particular case of the CS studied
here. The authors of [KR~96] have not realized this fact and, moreover, they write
in the Introduction: ``~\dots The coherent states thus obtained are different from
those defined in this paper (Ref. [DG~93]). Nevertheless, it seems to us that the
approach presented herein is a better one". These CS are defined as
$$
\ket{\xi} = \sum_j \xi^{-j} e^{-j^2/2}\, \ket{j},   \tag7.1
$$
where $\xi=e^{-l+i\phi}$, $l \in \R$, $\phi \in S^1$ and $\ket{j}$ are the
eigenvectors of the angular momentum operator. Two cases are considered in
[KR~96]: boson case when $j$ takes integer values, and fermion case when $j$
takes half-integer values.

In the following, we are going to prove that the CS (7.1) are particular 
cases of our CS $\ket{z^*;k}$. We have (see Section 4)
$$
\ket{z^*;k} = \sum_{n=-\infty}^\infty \p_n(z)^*\, \ket{n;k},  \tag7.2
$$
where $z = \om q -ip$, and after (4.7)
$$
\p_n(z)^* = \left(\frac{4\pi\hbar}{a^2\om}\right)^{1/4}
\exp\bigl(-\frac{\hbar}{2\om} (\frac{2\pi}a n + k)^2\bigr)\,
\exp\bigl(-\frac i\om (\frac{2\pi}a n + k)z^*\bigr).     \tag7.3
$$
By analogy with Ref. [KR~96], we set from now on $\hbar = 1$ and
$a = 2\pi$, so that $k \in [0,1)$. If we also put $\xi = \exp (iz^*/\om) =
\exp(-p/\om + iq)$, then (7.2) finally becomes
$$
\ket{z^*;k} = \left(\frac1{\pi\om}\right)^{1/4} \sum_{n=-\infty}^\infty
e^{-(n + k)^2/(2\om)}\, \xi^{-(n + k)}\, \ket{n;k}.     \tag7.4
$$
Now, simply comparing expressions (7.1) and (7.4) we see that both coincide 
(up to a constant factor) if we set $\omega = 1$ and $k = 0$ for the boson
case, or $k=\pi/a = 1/2$ for the fermion case. Indeed, for $k = 0$ we get
$$
\ket{z^*;0} = \left(\frac1\pi\right)^{1/4} \sum_{n=-\infty}^\infty
e^{-n^2/2}\, \xi^{-n}\, \ket{n;0},    \tag7.5
$$
which obviously coincides with (7.1) when $j$ takes integer values, because
expre\-ssion (6.15) shows that $\ket{n;0}$ are the boson eigenvectors of the
angular momentum operator. In the same way, for $k = 1/2$ we get
$$
\ket{z^*;\frac12} = \left(\frac1\pi\right)^{1/4} \sum_{n=-\infty}^\infty
e^{-(n + 1/2)^2/2}\, \xi^{-(n + 1/2)}\, \ket{n;\frac12},   \tag7.6
$$
that also equals (7.1) when $j$ takes half-integer values, since
$\ket{n;1/2}$ are now the fermion eigenvectors of the angular momentum 
operator, as we can see in expression (6.15). This ends the proof of our
statement.

From the study of the physical properties of these CS we can state
that they are very similar to the Heisenberg CS on $\R$, provided that
the wideness of the wave function is small in comparison with the
length of the configuration space $S^1$. Otherwise, the properties of
these CS drastically depend on the values of $p$. Moreover,
all the physical properties have a periodic behaviour in terms of $p$.

It is worthy to note that our CS are ``quasi-minimal", i.e., although
they do not minimize the Heisenberg uncertainty relation, the product of
the dispersions of the angle and momentum operators is upperly bounded
by $\hbar$.

Finally, we mention that these CS may be used to quantize the cylinder by 
means of the Weyl correspondence [DG~92] [DG~93] [GO~96]. Work in this
direction is in progress and the results will be published elsewhere.
\bigskip
\bigskip


\head Acknowledgments \endhead

We must acknowledge to Prof. Stephan de Bi\'evre  for his fruitful
commentaries.
This work has been partially supported by DGES of the  Ministerio de 
Educaci\'on y Cultura of Spain under Project PB95-0719, and the Junta de
Castilla y Le\'on (Spain).



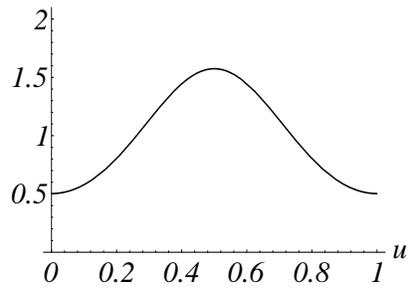
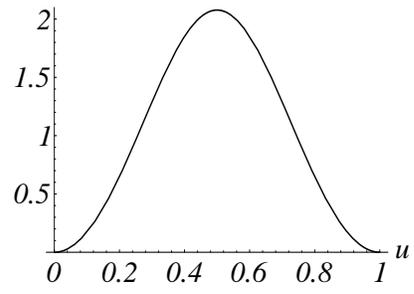
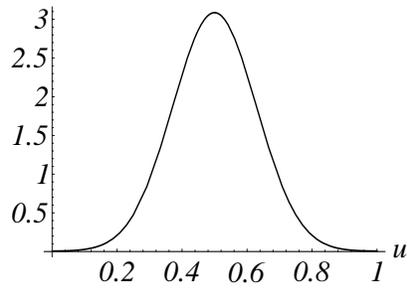
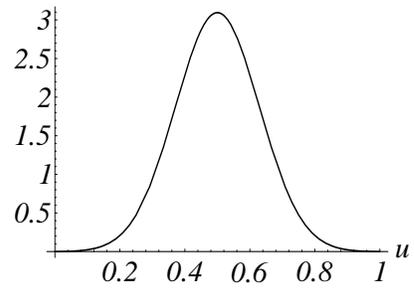
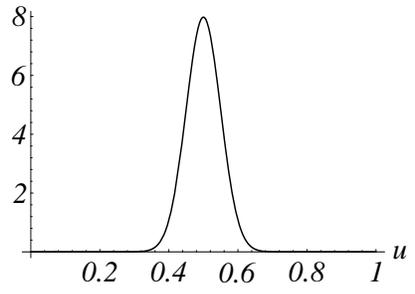
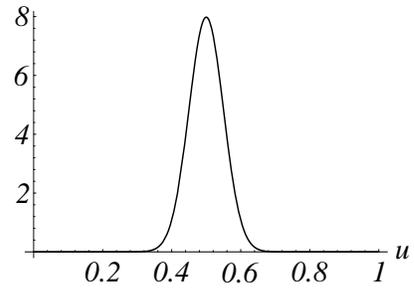
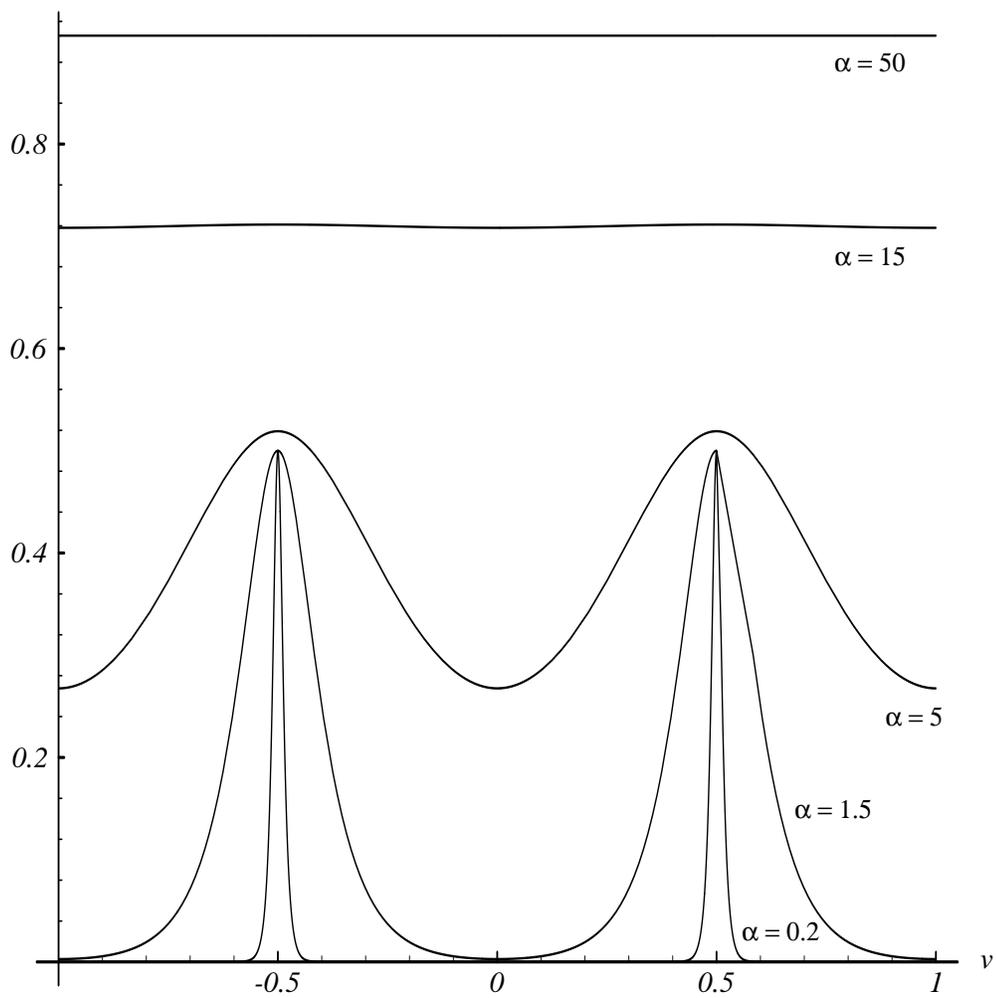
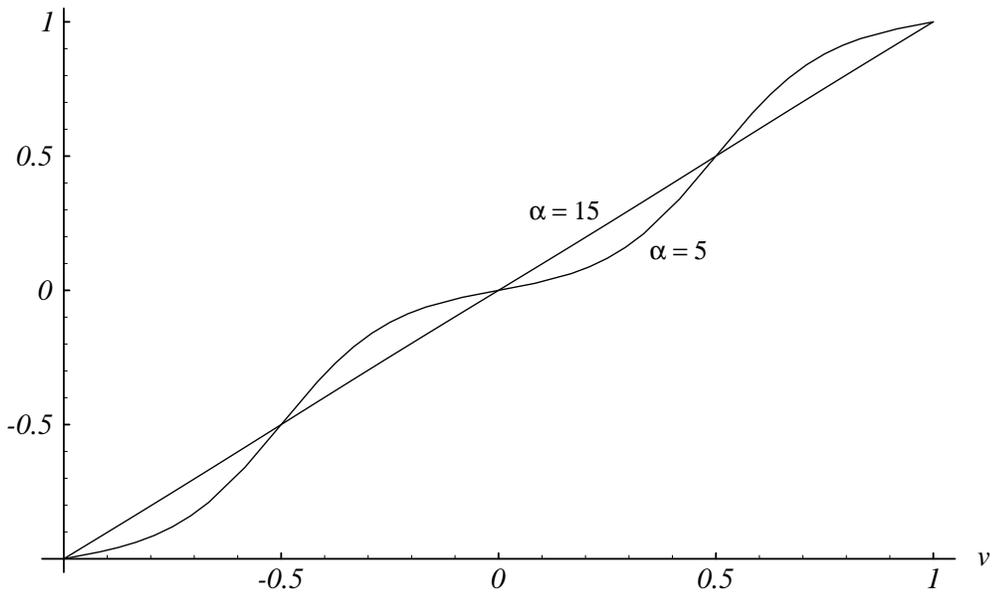
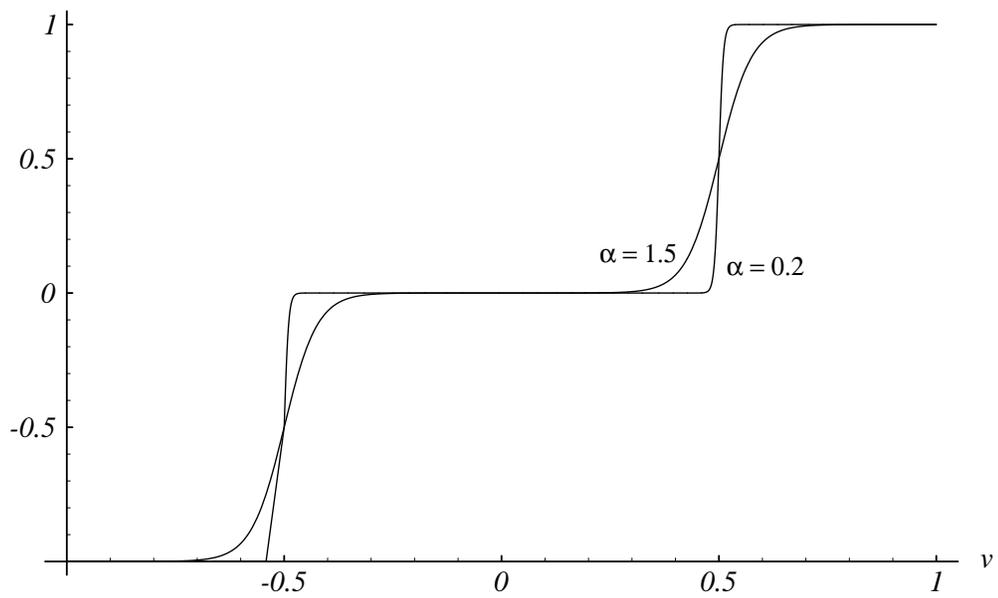
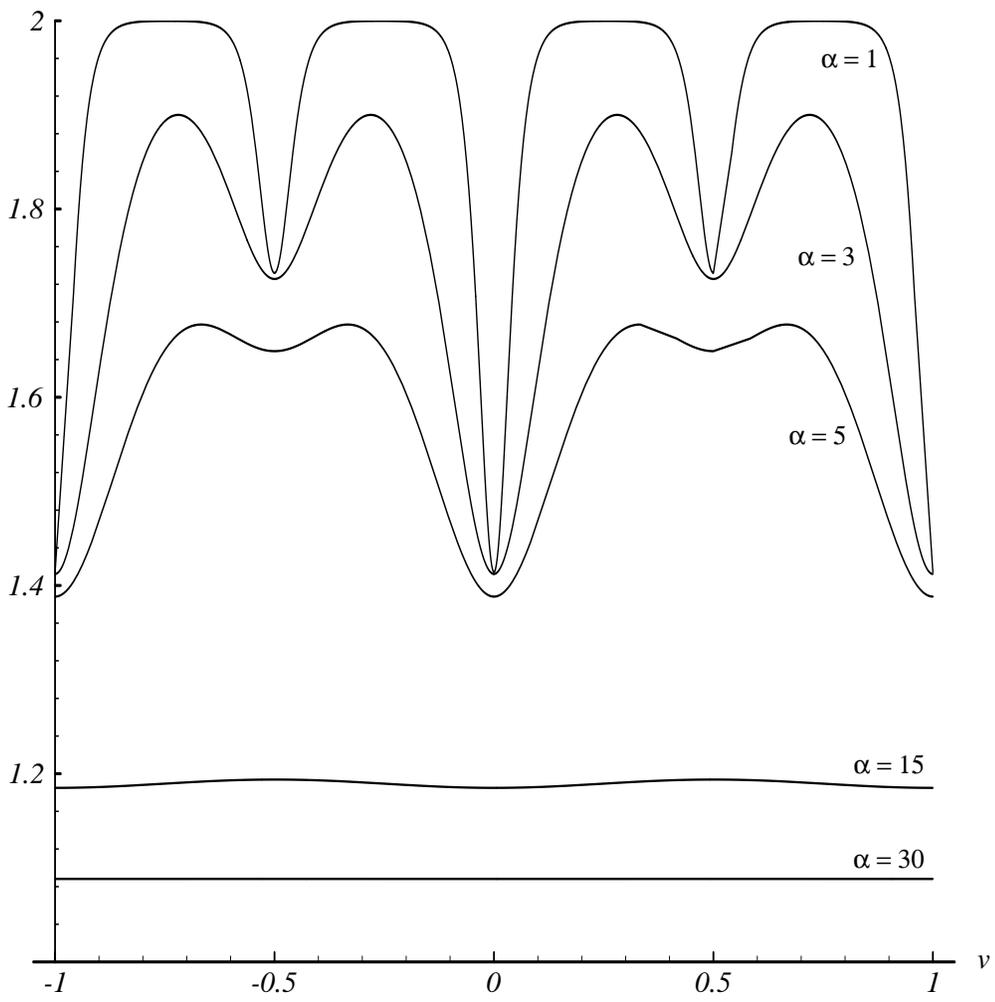
\end